\documentclass[conference]{IEEEtran}
\usepackage{amsmath,amssymb}
\usepackage{amsfonts}
\usepackage[dvips]{graphicx}
\usepackage{dsfont}
\usepackage{algorithmic}
\usepackage{balance}
\usepackage{algorithm}
\usepackage{mathtools}
\usepackage{comment}

\usepackage{color}
\usepackage{pgfplots}           %for plotting graphs
\pgfplotsset{compat=1.16}
\usepackage{caption}

\usepackage{enumerate}
\usepackage{graphics}
\usepackage{cite}
\usepackage{mathrsfs}
\usepackage{subcaption}
\usepackage{stfloats}
\usepackage{xfrac}

\usepackage{tikz}
\usepackage{mathdots}
\usepackage{yhmath}
\usepackage{cancel}
\usepackage{color}
\usepackage{siunitx}
\usepackage{array}
\usepackage{multirow}
\usepackage{textcomp}
\usepackage{gensymb}
\usepackage{tabularx}
\usepackage{booktabs}
\usetikzlibrary{fadings}
\usetikzlibrary{patterns}
\usetikzlibrary{shadows.blur}
\usepackage{graphicx}

\DeclarePairedDelimiter\ceil{\lceil}{\rceil}
\DeclarePairedDelimiter\floor{\lfloor}{\rfloor}
\usepackage{mathtools, nccmath}

\DeclarePairedDelimiter{\nint}\lfloor\rceil

\newtheorem{proposition}{Proposition}

\newtheorem{definition}{Definition}

\begin{document}

\title{\huge On the Performance of RIS-assisted Networks with HQAM}
%Advancing 6G Uplink with Non-Orthogonal Strategies \\ in Random Source Deployments}

\author{\IEEEauthorblockN{Thrassos K. Oikonomou\IEEEauthorrefmark{1}, Dimitrios Tyrovolas\IEEEauthorrefmark{1}, Sotiris A. Tegos\IEEEauthorrefmark{1}\IEEEauthorrefmark{2},\\ Panagiotis D. Diamantoulakis\IEEEauthorrefmark{1}, Panagiotis Sarigiannidis\IEEEauthorrefmark{2}, 
Christos Liaskos\IEEEauthorrefmark{3}, George K. Karagiannidis\IEEEauthorrefmark{1}\IEEEauthorrefmark{4}}

\IEEEauthorblockA{\IEEEauthorrefmark{1}Department of Electrical and Computer Engineering, Aristotle University of Thessaloniki, 54124 Thessaloniki, Greece}
\IEEEauthorblockA{e-mail: \{toikonom, tyrovolas, tegosoti, padiaman, geokarag\}@auth.gr}
\IEEEauthorblockA{\IEEEauthorrefmark{2}Department of Electrical and Computer Engineering, University of Western Macedonia, 50100, Kozani, Greece}
\IEEEauthorblockA{e-mail: psarigiannidis@uowm.gr}
\IEEEauthorblockA{\IEEEauthorrefmark{3}Computer Science Engineering Department, University of Ioannina, Ioannina, Greece. e-mail: cliaskos@ics.forth.gr}
\IEEEauthorblockA{\IEEEauthorrefmark{4}Artificial Intelligence \& Cyber Systems Research Center, Lebanese American University (LAU), Lebanon}
\vspace{-6mm}
}

\maketitle	
\begin{abstract}
In this paper, we investigate the application of hexagonal quadrature amplitude modulation (HQAM) in reconfigurable intelligent surface (RIS)-assisted networks, specifically focusing on its efficiency in reducing the number of required reflecting elements. Specifically, we present analytical expressions for the average symbol error probability (ASEP) and propose a new metric for conditioned energy efficiency, which assesses the network's energy consumption while ensuring the ASEP remains below a certain threshold. Additionally, we introduce an innovative detection algorithm for HQAM constellations that implements sphere decoding in $\mathcal{O}\left(1\right)$ complexity. Finally, our study reveals that HQAM significantly enhances both the ASEP and energy efficiency compared to traditional quadrature amplitude modulation (QAM) schemes.
\end{abstract}
%\vspace{-0.2cm}
\begin{IEEEkeywords}
    HQAM, Reconfigurable Intelligent Surfaces (RISs), Energy Efficiency, Average Symbol Error Probability, Detection Algorithm
\end{IEEEkeywords}

%\vspace{-0.3cm}
\section{Introduction}\label{S:Intro}
Motivated by the rise of innovative applications such as extended reality, smart industries, and autonomous vehicles, there is a growing demand to enhance the capabilities of wireless networks \cite{segata2023}. In this direction, a new wireless communication paradigm has emerged, named programmable wireless environment (PWE), aiming to transform the wireless propagation phenomenon into a software-defined process \cite{liaskos2018}. This transformation necessitates coating propagation environments with reconfigurable intelligent surfaces (RISs), which are planar structures capable of modifying the characteristics of the waves impinging upon them, including their direction and polarization \cite{direnzo-access}. In more detail, RISs can be appropriately configured to improve the quality of the wireless channels, as well as direct intelligently signal propagation to reduce power losses which is crucial for the sustainability of future 6G networks \cite{zeris}. Thus, RISs can ensure seamless connectivity for demanding futuristic applications but also enhance the network's energy efficiency, paving the way for ubiquitous ultra-reliable green communications.

To reliably perform various RIS functionalities, it is recognized that a significant number of reflecting elements is essential. However, considering that each element contributes to the network's power consumption through the tuning mechanism of their impedance, it becomes imperative to expand the capabilities of digital communication systems to ensure reliable performance with as few reflecting elements as possible. Within this context, hexagonal quadrature amplitude modulation (HQAM) stands out as a novel modulation scheme, due to its efficient and compact allocation of symbols on the 2D plane \cite{alouini2021, hqam_th}. Specifically, HQAM represents an appropriate modulation scheme for applications that prioritize energy savings, owing to its hexagonal lattice's ability to maximize the use of available space, thereby leading to a reduction in energy consumption. In this direction, by taking into account the advantages of HQAM, the authors of \cite{alouini2021} provided a comprehensive comparison of HQAM over existing QAM constellations and justify its supremacy in terms of symbol error rate and energy efficiency. Furthermore, the authors of \cite{hqam_th} provided a tight closed-form approximation for the symbol error probability of HQAM, as well as a detection algorithm for HQAM constellations of $\mathcal{O}\left(\mathrm{log} \sqrt{M} \right)$ complexity. Finally, \cite{alouini2020} investigated the effects of outdated channel state information, pointing errors, and atmospheric turbulence on the average symbol error probability (ASEP) for a mixed FSO/RF system that utilizes HQAM. However, to the best of the authors' knowledge, there exists no work that quantifies the effect of HQAM in an RIS-assisted network, particularly in terms of reducing the number of reflecting elements.

In this work, we analyze the performance of an RIS-assisted network that utilizes HQAM scheme, focusing on its potential to reduce the number of reflecting elements required for maintaining high-quality communication. Specifically, we derive analytical expressions for the ASEP and introduce a novel metric named conditioned energy efficiency, which evaluates the network energy efficiency while keeping the ASEP below a predefined threshold. Furthermore, we propose a novel detection algorithm for HQAM that implements sphere decoding in $\mathcal{O}\left(1\right)$ complexity, thereby enhancing the practical applicability of HQAM in future wireless communication systems. Finally, our simulation results reveal that HQAM offers significant improvements in the ASEP and the energy efficiency of RIS-assisted networks over traditional QAM schemes.

% The remainder of this paper is organized as follows. The system model is described in Section \ref{sysmodl}. The performance analysis of the considered network is presented in Section \ref{analysis} and the proposed detection algorithm is presented in Section \ref{S:detection}. Finally, numerical results are presented in Section \ref{secnum} and Section \ref{conclusion} concludes the paper.

\section{System Model}\label{sysmodl}
We consider an uplink communication network that consists of a single-antenna base station (BS) and a single antenna communication node (CN). Due to the harsh wireless propagation environment, it is assumed that there is no direct communication link available to facilitate the connection between the CN and BS. To enhance the received power at the BS, we employ a RIS with $N$ reflecting elements which assists the CN-BS communication by steering the CN transmissions towards the BS. Therefore, considering the RIS reflection path, the baseband equivalent of the received symbol at the BS can be expressed as
\begin{equation}\label{}
\small
    \begin{aligned}
        y = \sqrt{P_{t} G l_p}\sum_{i=1}^{N}\lvert h_{i1}\rvert\lvert h_{i2}\rvert e^{(\omega_{i} + \arg\{h_{i1}\} + \arg\{h_{i2}\})}s + w,
    \end{aligned}
\end{equation}
where $s$ is the transmitted symbol from a constellation $\mathcal{C}$ with unitary average enegy, i.e., $\mathbb{E}[\lvert E_{s}] \rvert = \mathbb{E}[\lvert s\rvert^2]=1$ where $\mathbb{E}[\cdot]$ denotes the expectation of a random variable (RV), while $\lvert \cdot\rvert$ and $\arg\{\cdot\}$ denote the magnitude and the argument of a complex number, respectively. Also, $P_{t}$ is the transmit power, $G=G_{t}G_{r}$ denotes the product of the CN and AP antenna gains, and $h_{1i}$ and $h_{2i}$ are the complex channel coefficients that correspond to the links between the CN and the $i$-th reflecting element and between the $i$-th reflecting element and the BS, respectively. More precisely, the RVs $\lvert h_{1i} \rvert$ and $\lvert h_{2i} \rvert$ are assumed to follow Nakagami-$m$ distribution with shape parameter $m$ and spread parameter $\Omega$, which can describe accurately realistic communication scenarios characterized by severe or light fading. Moreover, $w$ is the complex additive white Gaussian noise (AWGN) with zero mean and standard deviation $\sigma_{n}^2 = N_0$, $\omega_{i}$ is the phase correction term induced by the $i$-th element, and $l_p$ is the path loss corresponding to the CN-RIS-BS link. Specifically, $l_p$ can be modeled as $l_p = C_{0}^2\big(\frac{d_{0}}{d_{1}d_{2}}\big)^n$, where $n$ expresses the path loss exponent, $C_{0}$ denotes the path loss of CN-RIS and RIS-BS links at the reference distance $d_{0}$, while $d_{1}$ and $d_{2}$ denote the distances of the CN-RIS and RIS-BS links, respectively. 

To maximize the signal-to-noise ratio (SNR) at the receiver, the phase shift of the $i$-th reflecting element $\omega_{i}$ is ideally chosen to nullify the combined phase shift $\arg\{h_{i1}\} + \arg\{h_{i2}\}$ \cite{direnzo-access}. However, considering the reflecting elements' impedance being controlled by $q$ PIN diodes, there are $2^{q}$ distinct patterns of phase shifts for each element. Therefore, the received signal $y$ can be rewritten as
\begin{equation}
\small
    \begin{aligned}
        y = \sqrt{P_{t}G l_p} h s + w,
    \end{aligned}
\end{equation}
where $h=\sum_{i=1}^{N}\lvert h_{1i}\rvert\lvert h_{2i}\rvert e^{\phi_{i}}$, and $\phi_{i}$ is a uniformly distributed RV over $[-2^{-q}\pi, 2^{-q}\pi]$ \cite{hongliang}. Finally, by assuming perfect channel state information (CSI) knowledge, the received symbol $r$ can be obtained by performing channel inversion.

%\subsection{HQAM Scheme}
%Driven by the emergence of RIS-assisted networks, it is of paramount importance to offer high QoS in a cost and energy-efficient manner.  In this context, HQAM is positioned as a critical modulation technique capable of playing a pivotal role in future wireless systems, primarily due to its compact allocation of symbols on the 2D plane. In more detail, an $M$-ary HQAM constellation, as illustrated in Fig. \ref{fig:hqam}, is defined by a set $\mathcal{C}_{H} = \{ \mathbf{s_{z}}\in \mathbb{R}^2, z=0,...,M-1\}$ of $M$ symbols, where $\mathcal{C}_{H}$ is a subset of infinite grid $S = \{\mathbf{v}\in \mathbb{R}^2: \mathbf{v}=c_{1}\mathbf{v_{1}}+c_{2}\mathbf{v_{2}}+\mathbf{x_{0}}\}$, $\mathbf{v_{1}}=[\sfrac{d_{\mathrm{min}}}{2},0]^T$ and $\mathbf{v_{2}}=[0,\sfrac{\sqrt{3}d_{\mathrm{min}}}{2}]^T$ are the basis vectors of the 2D grid, $d_{\mathrm{min}}$ is the minimum distance between symbols, $c_{1}, c_{2}\in \mathbb{Z}$ and $\mathbf{x_{0}}\in \mathbb{R}^2$ is the offset of the grid. Finally, regarding the maximum-likelihood decision regions, the internal symbols are equilateral hexagons, with angles $\sfrac{2\pi}{3}$, while the infinite region of the external symbols has one or more angles equal to $\sfrac{2\pi}{3}$. 

\begin{figure}
\centering
\begin{tikzpicture}[scale=0.5]
\begin{axis}[
        xtick=\empty,
        ytick=\empty,
        xlabel={$\text{In-Phase}$},
        ylabel={$\text{Quadrature}$},
        xlabel style={below},
        ylabel style={above},
        xmin=-1.5,
        xmax=1.5,
        ymin=-1.5,
        ymax=1.5,
    ]
    \addplot [only marks, black] table {figures/hqam_iq_64.txt};
    \addplot [no markers, update limits=false] table {figures/hqam_voronoi_64.txt};
\end{axis}
\end{tikzpicture}
\caption{64-HQAM constellation}
\label{fig:hqam}
\end{figure}
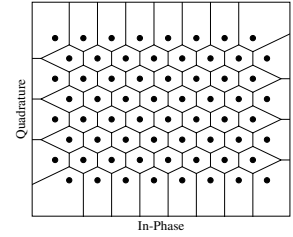

\section{Performance Analysis}\label{analysis}
In this section, we extract analytical expressions for the examined network’s ASEP and introduce a new metric named conditioned energy efficiency, which quantifies the ratio of the throughput to the network's energy consumption under the condition that ASEP remains below a predefined threshold.

The Symbol Error Probability (SEP) stands as one of the most crucial metrics for assessing the performance of a modulation scheme. However, due to HQAM hexagonal lattices as shown in Fig. \ref{fig:hqam}, evaluating the exact SEP is challenging. Therefore, by approximating the area of the regular hexagons with a circle as shown in \cite{hqam_th}, the SEP can be tightly approximated as
\begin{equation}\label{eq:sep-hqam-awgn}
\small
    \begin{aligned}
        P_{s}(\gamma_{r}, k_c) & \approx  \dfrac{2M-b}{2M} e^{-\gamma_{r} B} 
         + \dfrac{b}{M} Q\left(\sqrt{\gamma_{r}}A\right),
    \end{aligned}
\end{equation}
where $Q(x) = \frac{1}{\sqrt{2\pi}} \int_x^\infty \exp\left(-\frac{u^2}{2}\right) \, du$ is the Gaussian $Q$-function, $\gamma_{r}$ is the received SNR, $B = \frac{d_{\mathrm{min}}^2 k_c^2}{3E_s}+\frac{d_{\mathrm{min}}^2k_c(1-k_c)}{\sqrt{3}E_s}+\frac{d_{\mathrm{min}}^2(1-k_c)^2}{4E_s}$ and $A = \sqrt{\frac{2d_{\mathrm{min}}^2k_c^2}{3E_s}} + \sqrt{\frac{d_{\mathrm{min}}^2(1-k_c)^2}{2E_s}}$, respectively. Moreover, $d_{\mathrm{min}}$ denotes the nearest-neighbour distance and is expressed as
\begin{equation}
\small
    d_{\mathrm{min}} = \sqrt{\frac{12E_s}{7M-4}},
\end{equation}
and $b$ denotes the number of the constellation's external symbols \cite{hqam_th}. Finally, by taking into account the constellation order $M$, the values of $k_c$ is given in Table I. However, considering the existence of fading channels in the examined scenario, it is imperative to derive the probability density function (PDF) $f_{\vert h\vert}(\cdot)$ of $\vert h\vert$, to derive the ASEP of the considered RIS-assisted network. To this end, we provide a closed-form expression that tightly approximates the PDF of the considered channel $h$.

\begin{table}
    \begin{center}
        \caption{Parameter $k_c$ for $M$-HQAM.}
        \label{rho}
        \begin{tabular}{|c|c|c|c|c|c|c|c|}
            \hline
            $M$ &$16$&$64$&$256$&$1024$ \\ \hline% 0.65416708 \\ 
            $k$ &0.8711505&0.5222431&0.3936315&0.2982858\\ \hline
        \end{tabular}
    \end{center}
\end{table}

\begin{proposition}
    The PDF of $\lvert h \rvert$ for the considered RIS-assisted network can be tightly approximated as
    \begin{equation}\label{eq:h-approx}
    \small
        f_{\vert h \vert}(x) \approx \frac{2 {m_t}^{m_t}}{\Gamma(m_t){\Omega_t}^{m_t}} x^{2 m_t -1} e^{-\frac{m_t x^2}{\Omega_t}},
    \end{equation}
where $\Gamma(\cdot)$ is the gamma function, $m_{t} = \frac{I_{1}^2 }{I_{2} - I_{1}^2}$, $\Omega_{t} = I_{1}^2 $ and $I_1$, $I_2$ are given by \eqref{eq:I1} and \eqref{eq:I2} at the top of the next page, respectively.
\end{proposition}

\begin{figure*}[t]
    \begin{equation}\label{eq:I1}
    \small
        \begin{aligned}
             I_1 = N \left( \Omega^2 + (N-1)\left(\frac{2^{q}\Omega \sin\left(\frac{\pi}{2^q}\right)\Gamma\left(m+\frac{1}{2}\right)}{m\pi\Gamma\left(m\right)}\right)^2 \right) 
        \end{aligned}
    \end{equation}
\hrule
    \begin{equation}\label{eq:I2}
    \small
        \begin{aligned}
            I_2 = 
            &N(N-1)(N-2)(N-3) \left(\frac{\Gamma\left(m + \frac{1}{2}\right)}{\Gamma(m)}\right)^8\frac{\Omega^4}{m^4} \frac{2^{4q}}{\pi^4} \sin^4\left(\frac{\pi}{2^q}\right) + 4N(N-1)\left(\frac{\Gamma\left(m + \frac{3}{2}\right)}{\Gamma(m)}\right)^2\frac{\Omega^3}{m^3}\frac{4^q}{\pi^2}\sin^2\left(\frac{\pi}{2^q}\right)\\
            &+ N(N-1)(N-2) \Omega^2 \left(\frac{\Gamma(m + \frac{1}{2})}{\Gamma(m)}\right)^4\frac{\Omega^2}{m^2} \left( \frac{2^{2q+1}}{\pi^2}\sin^2\left(\frac{\pi}{2^q}\right) +   4\left(\frac{1}{2}+\frac{2^{q-2}}{\pi}\sin\left(\frac{\pi}{2^{q-1}}\right) \right)\frac{4^q}{\pi^2}\sin^2\left(\frac{\pi}{2^q}\right) \right) \\
            &+ N\left(\frac{\Gamma(m + 2)}{\Gamma(m)}\right)^2\frac{\Omega^4}{m^4} + N(N-1)\left( \Omega^4 + \Omega^4\left( 1 + \frac{2^{2q-2}}{\pi^2}\sin^2\left(\frac{\pi}{2^{q-1}}\right) \right)  \right)
        \end{aligned}
    \end{equation}
\hrule
\end{figure*}

\begin{IEEEproof}
Considering that $N$ is large, by applying the moment-matching technique \cite{tyr-harq}, $\lvert h \rvert ^2$ can be approximated as a gamma-distributed RV with scale parameter $k_t=\frac{I_{1}^2 }{I_{2} - I_{1}^2}$ and shape parameter $\theta_t=\frac{I_{2} - I_{1}^2}{I_{1}}$, where $I_1$ and $I_2$ are the first and the second moment of $\lvert h \rvert^2$, hence $\vert h\vert$ can be approximated as a Nakagami-$m$ RV with shape parameter $m_t=k_t$ and scale parameter $\Omega_t=k_t\theta_t$. Therefore, to tightly approximate $f_{\vert h \vert}(x)$, we need to calculate $I_1$ and $I_2$, which are equal to $\mathbb{E}[\lvert h \rvert ^2]$, and $\mathbb{E}[\lvert h \rvert ^4]$, respectively. Initially, after some algebraic manipulations, $\lvert h \rvert ^2$ and $\lvert h \rvert ^4$ can be expressed respectively as
\begin{equation}\label{h2}
\small
    \lvert h \rvert ^2 = \sum_{i=1}^{N} \sum_{j=1}^{N}\lvert h_{1i}\rvert\lvert h_{2i}\rvert \lvert h_{1j}\rvert\lvert h_{2j}\rvert \mathrm{cos}(\phi_i - \phi_j),
\end{equation}
and
\begin{equation}\label{h4}
\begin{aligned}
\small
    \lvert h \rvert ^4 = \sum_{i=1}^{N} \sum_{j=1}^{N} \sum_{k=1}^{N} \sum_{l=1}^{N}\lvert h_{1i}\rvert\lvert h_{2i}\rvert \lvert h_{1j}\rvert\lvert h_{2j}\rvert \lvert h_{1k}\rvert\lvert h_{2k}\rvert 
    \\ \times \lvert h_{1l}\rvert\lvert h_{2l}\rvert \mathrm{cos}(\phi_i - \phi_j)\mathrm{cos}(\phi_k - \phi_l).
  % \end{split}
\end{aligned}
\end{equation}
By observing \eqref{h2} and \eqref{h4}, it becomes evident that $I_1$ and $I_2$ will consist of $N^2$ and $N^4$ terms, respectively. In more detail, considering that $\mathbb{E}[\cdot]$ is a linear operator, $\mathbb{E}[\lvert h \rvert ^2]$ and $\mathbb{E}[\lvert h \rvert ^4]$ are equal to the sum of the mean values of each term of the summation. To this end, by identifying the expression of each term and calculating its mean value, we can obtain $I_1$ and $I_2$. In this direction, for the case of $I_1$,  the summation terms are the following:
\begin{itemize}
\small
    \item $N^2-N$ terms: $\lvert h_{1i}\rvert\lvert h_{2i}\rvert \lvert h_{1j}\rvert\lvert h_{2j}\rvert \mathrm{cos}(\phi_i - \phi_j)$, if $i\neq j$,
    \item $N$ terms: $\lvert h_{1i}\rvert ^2\lvert h_{2i}\rvert ^2$, if $i=j$.
\end{itemize}
Therefore, $I_1$ can be expressed as
\begin{equation}
\small
    \begin{split}
        I_1 = &\left(N^2-N\right)\mathbb{E}\Big[\lvert h_{1i}\rvert\lvert h_{2i}\rvert \lvert h_{1j}\rvert\lvert h_{2j}\rvert \mathrm{cos}(\phi_i - \phi_j)\Big] \\
        & + N \mathbb{E}\Big[\lvert h_{1i}\rvert ^2\lvert h_{2i}\rvert ^2 \Big].
    \end{split}
\end{equation}
 By taking into account that $\lvert h_{1i}\rvert,\lvert h_{2i}\rvert, \lvert h_{1j}\rvert, \lvert h_{2j}\rvert,$ and $\mathrm{cos}(\phi_i - \phi_j)$ are independent RVs with each other, $I_1$ can be rewritten as
\begin{equation}
\small
\begin{split}
    I_1= &\left(N^2-N\right)\mathbb{E}\Big[\lvert h_{1i}\rvert\Big] \mathbb{E}\Big[\lvert h_{2i}\rvert\Big] \mathbb{E}\Big[\lvert h_{1j}\rvert\Big] \\
        & \times \mathbb{E}\Big[\lvert h_{2j}\rvert\Big] \mathbb{E}\Big[\mathrm{cos}(\phi_i - \phi_j)\Big] + N \mathbb{E}\Big[\lvert h_{1i}\rvert^2\Big] \mathbb{E}\Big[\lvert h_{2i}\rvert^2\Big]. 
\end{split}
\end{equation}
Moreover, considering that $\lvert h_{1i}\rvert$ and $\lvert h_{2i}\rvert$ are Nakagami-$m$ distributed RVs and that $\phi$ is a uniformly distributed RV over $[-2^{-q}\pi, 2^{-q}\pi]$, after some algebraic manipulations, we derive that $ \mathbb{E}\Big[\lvert h_{1i}\rvert\Big]=\mathbb{E}\Big[\lvert h_{2i}\rvert\Big]=\frac{\Gamma(m+1/2)}{\Gamma(m)} \left( \frac{\Omega}{m}\right)^{\frac{1}{2}}$, $\mathbb{E}\Big[\lvert h_{1j}\rvert^2\Big]=\mathbb{E}\Big[\lvert h_{2j}\rvert^2\Big]=\frac{\Gamma(m+1)}{\Gamma(m)} \left( \frac{\Omega}{m}\right)$ and $\mathbb{E}\Big[\mathrm{cos}(\phi_i - \phi_j)\Big]= \frac{4^q}{\pi}\mathrm{sin}^2\left(\frac{\pi}{2^q}\right)$, thus we can calculate $I_1$ as in \eqref{eq:I1}. Similarly, by identifying the different terms of \eqref{h4}, we can obtain \eqref{eq:I2}, thus enabling the calculation of $m_t$ and $\Omega_t$, which concludes the proof.
\end{IEEEproof}

Next, we provide the ASEP for the considered RIS-assisted network for the case where $s$ belongs to an $M$-ary HQAM constellation.
\begin{proposition}
    For the case where $s$ belongs to an $M$-ary HQAM, the ASEP of the considered RIS-assisted network can be tightly approximated as
    \begin{equation} \label{eq:ASEP}
    \small
            P_{a} \approx \frac{2M-b}{2M} \bigg( \frac{\tilde{m_{t}}}{\tilde{m_{t}} + \frac{B}{\Omega_{t}}\bar{\gamma}}\bigg)^{\tilde{m_{t}}} + \frac{b}{M} P_{1},
    \end{equation}
    where $\bar{\gamma} = \frac{P_t G l_p}{\sigma_{n}^2}$, $\tilde{m_t} = \nint{m_t}$, where $\nint{\cdot}$ is the round function, and $P_1$ is given as
    \begin{equation}
    \small
        \begin{aligned}
            P_1 = \frac{1}{2} -\frac{1}{2}\sum_{z=0}^{m_{t}-1}\frac{{m_{t}}^{z}\sqrt{A^2\bar{\gamma}}(2z)!}{ ({4\Omega_{t}})^{z}\sqrt{A^2\bar{\gamma} + \frac{m_{t}}{\Omega_{t}}}  
            (\frac{A^2\bar{\gamma}}{2}+\frac{m_{t}}{\Omega_{t}})^z z!}.
        \end{aligned}
    \end{equation}

\end{proposition}
\begin{IEEEproof}
The ASEP for the considered RIS-assisted network for the case where $M$-HQAM is utilized is written as 
\begin{equation}  \label{eq:asep-integral}
\small
    \begin{aligned}
        P_{a} = \int_{0}^{\infty} P_{s}(x^2 \bar{\gamma}, k) f_{x}(x) \,dx.
    \end{aligned}
\end{equation}
By substituting \eqref{eq:sep-hqam-awgn} and \eqref{eq:h-approx} in \eqref{eq:asep-integral}, we obtain that 
\begin{equation}  \label{eq:ASEP-hqam}
\small
    \begin{aligned}
        P_{a} &= \frac{ (2M-b){m_t}^{m_t}}{M\Gamma(m_t){\Omega_t}^{m_t}}\int_{0}^{\infty} \  e^{-x^2( \bar{\gamma} B +\frac{m_t}{\Omega_t})} x^{2 m_t -1} \,dx, \\
         &+ \frac{2 b{m_t}^{m_t}}{M\Gamma(m_t){\Omega_t}^{m_t}} \int_{0}^{\infty} Q\left(x\sqrt{\bar{\gamma}}A\right) x^{2 m_t -1} e^{-\frac{m_t x^2}{\Omega_t}}\,dx.
    \end{aligned}
\end{equation}
By utilizing the equation $Q(x) = \frac{1}{2} - \frac{1}{2}\mathrm{erf}(\frac{x}{\sqrt{2}})$, where $\mathrm{erf}(\cdot)$ is the error function, and approximating $m_t$ with $\nint{m_t}$ the integrals in \eqref{eq:ASEP-hqam} can be calculated as in \cite[(3.461/3)]{rhyzik} and \cite[(2.6.2/1)]{korotkov2020integrals}, which concludes the proof.
\end{IEEEproof}

Next, we define the conditioned energy efficiency metric, which quantifies the network's energy efficiency under the condition that ASEP remains below a predefined threshold.
\begin{definition}
 The conditioned energy efficiency of a wireless communication system that utilizes an $M$-ary constellation is defined as the ratio of the network's throughput to the network's energy consumption, under the condition that ASEP remains below a predefined threshold, and can be expressed as
 \begin{equation}\label{eq:CondEE}
 \small
     E_{c} = U(P_{f}-P_{v})\frac{\left(1- P_{f}\right) B \mathrm{log}_2 \left(M\right)}{P_c},
 \end{equation}
where $U(\cdot)$ is the unit step function defined as $U\left(x\right)=1$, if $x\geq0$ and $U\left(x\right)=0$, otherwise,  $P_f$ and $B$ are the network's ASEP and bandwidth, respectively, $P_{v}$ is the ASEP threshold, and $P_c$ is the network's power consumption.
\end{definition}

Considering the above definition, by substituting \eqref{eq:ASEP} in \eqref{eq:CondEE}, the conditioned energy efficiency for the examined communication scenario is given as
\begin{equation}
\small
    E_{c} = U(P_{a}-P_{v})\frac{\left(1- P_{a}\right) B \mathrm{log}_2 \left(M\right)}{P_t + P_{\mathrm{ctr}}+ q N P_{\mathrm{PIN}}},
\end{equation}
where $P_{\mathrm{ctr}}$ is the power consumption of the RIS controller, and $P_{\mathrm{PIN}}$ is the power consumption of each PIN diode.

\section{Detection Scheme}\label{S:detection}
Incorporating HQAM into communication systems can significantly boost the system's energy efficiency.  However, current research highlights that the detection algorithms for HQAM are more complex compared to those for traditional constellations (e.g., QAM), leading to a less favorable performance of HQAM in real-world applications. To address this issue, in this section, we introduce an innovative detection algorithm tailored for HQAM constellations that achieves a computational complexity of $\mathcal{O}(1)$, while maintaining performance levels comparable to the optimal maximum likelihood detection (MLD).
	
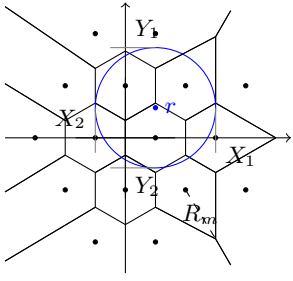
\begin{figure}
    \centering
    \newcommand{\polygon}[2]{%
        let \n{len} = {2*#2*tan(360/(2*#1))} in
        ++(0,-#2) ++(\n{len}/2,0) \foreach \x in {1,...,#1} { -- ++(\x*360/#1:\n{len})}}
    \begin{tikzpicture}[scale=0.4]
        \tikzstyle{every node}=[font=\small]
        
        %draw cartesian plane
        %\draw[step=0.5, gray, very thin] (-2,-2) grid (2.0,2.0);
        %draw x,y axes
        \draw[->] (-4,0) -- (5.5,0) coordinate (x axis);
        \draw[->] (0,-4.5) -- (0,4.5) coordinate (y axis);

        %draw reg hexagonal decision regions
        \draw[rotate around={90:(1,0)}] (1,0) \polygon{6}{1};
        \draw[rotate around={90:(-1,0)}] (-1,0) \polygon{6}{1};
        \draw[rotate around={90:(0,-1.73205081)}] (0,-1.73205081) \polygon{6}{1};
        \draw[rotate around={90:(0,1.73205081)}] (0,1.73205081) \polygon{6}{1};
        %symbols
        \filldraw[black] (1,0) circle (2pt);
        \filldraw[black] (-1,0) circle (2pt);
        \filldraw[black] (0,-1.73205081) circle (2pt);
        \filldraw[black] (0,1.73205081) circle (2pt);
        \filldraw[black] (2,1.73205081) circle (2pt);
        \filldraw[black] (-2,1.73205081) circle (2pt);
        \filldraw[black] (-2,-1.73205081) circle (2pt);
        \filldraw[black] (2,-1.73205081) circle (2pt);
        \filldraw[black] (3,0) circle (2pt);
        \filldraw[black] (-3,0) circle (2pt);
        \filldraw[black] (-1,-3.46410162) circle (2pt);
        \filldraw[black] (1,-3.46410162) circle (2pt);
        \filldraw[black] (-1,3.46410162) circle (2pt);
        \filldraw[black] (1,3.46410162) circle (2pt);
        \filldraw[black] (4,1.73205081) circle (2pt);
        \filldraw[black] (4,-1.73205081) circle (2pt);
        %draw non hexagonal decision regions
        \filldraw[black] (2,0.57735) -- (3, 1.1546);
        \filldraw[black] (3, 1.1546) -- (3, 3.3641);
        \filldraw[black] (1, 2.3094) -- (3, 3.3641);
        
        \filldraw[black] (3, 1.1546)--(5,0);
        \filldraw[black] (2,-0.57735) -- (3, -1.1546);
        \filldraw[black] (3, -1.1546)--(5,0);
        \filldraw[black] (3, -1.1546)--(3,-3.3641);
        \filldraw[black] (1, -2.3094) -- (3, -3.3641);
        
        \filldraw[black] (-2,0.57735) -- (-4, 1.8);
        \filldraw[black] (-1, 2.3094) -- (-4, 4.3641);
        
        \filldraw[black] (-2,-0.57735) -- (-4, -1.8);
        \filldraw[black] (-1, -2.3094) -- (-4, -4.1);
        \filldraw[black] (3, 3.3641) -- (3.5, 4.230125);
        \filldraw[black] (3, -3.3641) -- (3.5, -4.230125);
        %\filldraw[black] (0, -2.88675) -- (0, -5);
        
        %R_{m} radius of detection
        \draw[black, dashed]  (2,-1.73205081)-- (3, -3.3641) node [midway] {$R_{m}$};
        
        %draw received symbol and circle of radius d_{min}/2
        \filldraw[blue] (1,1) circle (2pt) node[anchor=west] {$r$};
        \draw[blue] (1,1) circle (2);
        %draw regions of intersection with y and x axis of tangent lines in circle of r
        \draw[gray] (1,3) coordinate (A) -- (-0.5, 3) coordinate (B);
        \draw (0,0) coordinate (C) -- (0, 2) coordinate (D);
        \node[anchor=south west] at (intersection of  A--B and C--D){$Y_{1}$};
        \draw[gray] (1,-1) coordinate (E) -- (-0.5, -1) coordinate (F);
        \draw (0,0) coordinate (G) -- (0, -2) coordinate (H);
        \node[anchor=north west] at (intersection of  E--F and G--H){$Y_{2}$};
        
        \draw[gray] (3,1) coordinate (I) -- (3, -0.5) coordinate (J);
        \draw (0,0) coordinate (K) -- (1.6547, 0) coordinate (L);
        \node[anchor=north west] at (intersection of  I--J and K--L){$X_{1}$};
        \draw[gray] (-1,1) coordinate (M) -- (-1, -0.5) coordinate (N);
        \draw (0,0) coordinate (O) -- (-1.6547, 0) coordinate (P);
        \node[anchor=south east] at (intersection of  M--N and O--P){$X_{2}$};
        
    \end{tikzpicture}
    \caption{Detection scheme of HQAM}
    \label{det_scheme}
\end{figure}

The key idea of our algorithm lies in identifying potential constellation symbols close to the received symbol $r = x_{\mathrm{r}} + jy_{\mathrm{r}}$, thereby reducing the necessary number of computed Euclidean distances between $r$ and the constellation symbols. Within this framework, as in sphere decoding \cite{sphere-decoding}, we etch a circle centered on $r$ with a radius of $R_{m} = d_{\mathrm{min}}$ and then define the regions $R_{1} = [X_1, X_2]$ and $R_{2} = [Y_1, Y_2]$, where $X_1$, $X_2$, $Y_1$, and $Y_2$ are the tangent lines of the circle which are parallel to the axes, as illustrated in Fig. \ref{det_scheme}. The detection process is completed by obtaining the symbol $\hat{s} \in G = R_{1} \cap R_{2}$ with the minimum Euclidean distance from $r$. To accomplish this, it becomes evident that we need to identify the constellation symbols that are contained within the region $G$. In this direction, we store all the real values $x_{i}$ of the constellation symbols in ascending order within set $S_{x}$. Afterwards, we construct $A_{x_i}$, a two-column adjacent matrix for $x_{i}$, which is sorted in ascending order with respect to the values of its second column. In more detail, the first column enlists the constellation symbols with a real part equal to $x_i$, and the second column denotes their corresponding imaginary parts, and it is described as
\begin{equation}
\small
    A_{x_i} = \begin{bmatrix} 
        s_{i,1} & \operatorname{Im(s_{i,1})} \\
        \vdots & \vdots & \\
        s_{i,p} & \operatorname{Im(s_{i,p})} 
    \end{bmatrix},
\end{equation}
where $p$ denotes the number of symbols that their real part equals to $x_i$.

To design an $\mathcal{O}(1)$ detection algorithm tailored for HQAM constellations, it is essential to exploit their structure and the arrays $S_x$ and $A_x$.  In this direction, by noting that for an HQAM constellation the values in $S_x$ and the values in the second column of $A_x$  increase with a constant rate, we can formulate a linear interpolation function $f(\cdot)$ that returns the position of its input in the arrays, and thus, find the constellation symbols within $G$ in $\mathcal{O}(1)$ time complexity. In particular, when the input value $u$ of $f$ matches one of the array elements, $f$ returns an integer value representing the position of $u$ in the array. Conversely, if $u$ is not an array element, $f$ yields a non-integer value, where rounding this value results in the position of $S_x$ with the closest corresponding value to $u$. Therefore, given that the elements of $S_{x}$ are equispaced by $\frac{d_{\mathrm{min}}}{2}$ due to the HQAM geometry, we can define the linear interpolation function $f_{\chi}: \mathbb{R} \rightarrow \mathbb{R}$, which is given by 
\begin{equation}\label{inter_x}
\small
    \begin{aligned}
        f_{\chi}(u) = \frac{2}{d_{\mathrm{min}}}u + 1 - \frac{2x_1}{d_{\mathrm{min}}},
    \end{aligned}
\end{equation}
where $x_1$ is the first element of $S_x$. To this end, by utilizing \eqref{inter_x}, we can obtain the positions of the different $x_i\in S_x$ that satisfy $X_1 \leq x_{i} \leq X_2$, which are given as
\begin{equation} \label{eq:xi}
\small
    \begin{aligned}
        \ceil*{f_{\chi}(X_{1})}\leq i \leq \floor*{f_{\chi}(X_{2})},
    \end{aligned}
\end{equation}
where $\ceil*{\cdot}$ and $\floor*{\cdot}$ are the ceil and floor functions, respectively. In a similar manner, given that the elements of $A_{x_i}$ are equispaced by $\frac{\sqrt{3}d_{\mathrm{min}}}{2}$ due to the HQAM geometry, we can also define the linear interpolation function $f_{\psi,i}: \mathbb{R} \rightarrow \mathbb{R}$, which is given by
\begin{equation}\label{inter_Ax}
\small
    \begin{aligned}
        f_{\psi,i}(v) = \frac{2}{\sqrt{3}d_{\mathrm{min}}}v+1-\frac{2A_{x_i}(1,2)}{\sqrt{3}d_{\mathrm{min}}},
    \end{aligned}
\end{equation}
where $A_{x_i}(1,2)$ is the first element of the second column of the 2D matrix. Therefore, the $j$-th element of the second column of $A_{x_i}$ that satisfies $Y_1 \leq \operatorname{Im(s_{i,j})} \leq Y_2$ can be obtained as \color{black}
\begin{equation} \label{eq:j}
\small
    \begin{aligned}
        \ceil*{f_{\psi,i}(Y_1)}\leq j \leq \floor*{f_{\psi,i}(Y_2)}.
    \end{aligned}
\end{equation}
Consequently, we can obtain the set $S_c$ that contains the symbols within $G$ which can be expressed as 
\begin{equation}
\small
    \begin{aligned}
        S_c = \{s : s = A_{x_i}(j,1)\},
    \end{aligned}
\end{equation}
where $A_{x_i}(j,1)$ is the $j$-th element of the first column of $A_{x_i}$. Finally, the symbol that the proposed detector determines is obtained by
\begin{equation}
\small
    \begin{aligned}
        \hat{s} = \arg \min_{s\in S_c}\hspace{4px}\lvert r-s\rvert^2.
    \end{aligned}
\end{equation}
It should be highlighted that $\vert S \vert \leq 6$ for any $M$-ary HQAM constellation, thus, instead of comparing $M$ symbols, as the conventional MLD describes, we only need to calculate 6 Euclidean distances at most for any HQAM constellation.
	
To ensure that the aforementioned algorithm operates properly the region $G$ must contain at least one symbol. However, under low-SNR conditions, it becomes quite possible that the received symbol is located far away from any other constellation symbol, thus, resulting in an empty region $G$. To tackle this, we initialize arrays $Q_{e}$, $e=\{1,2,3,4\}$, with the external symbols of the constellation. These symbols are chosen such that at least a portion of their infinite decision region falls within the $e$-th quadrant of the 2D plane.
Subsequently, when the received symbol $r$ is situated in the $e$-th quadrant, we make a decision by selecting the symbol $z$ from array $Q_{e}$ with the smallest Euclidean distance to $r$.
The time complexity of linearly traversing array $Q_{e}$
to locate symbol $z$ is $\mathcal{O}(\sqrt{M})$.

% To address this, the arrays $Q_{i}$, $i=\{1,2,3,4\}$, are initialized containing the constellation's external symbols for which at least one part of their infinite decision region is located within the $i$-th quadrant of the 2D plane. Then, if the received symbol $r$ lies in the $i$-th quadrant, the decision is made by obtaining the symbol $z$ of the array $Q_{i}$ with the minimum Euclidean distance from $r$. The time complexity of linearly traversing array $Q_{i}$ to find symbol $z$ is $\mathcal{O}(\sqrt{M})$.

\begin{algorithm}
    \caption{Detection algorithm}
    \begin{algorithmic}[1]
        \raggedright
        \renewcommand{\algorithmicrequire}{\textbf{Input:}}
        \renewcommand{\algorithmicensure}{\textbf{Output:}}
        \REQUIRE Coordinates of the received symbol $r$
        \ENSURE  Detected symbol
        %\textit{Initialize}: Constellation %
        \STATE Etch a circle with center $r$ and radius $R_{m}$.  
        \STATE Calculate points $Y_{1}$, $Y_{2}$.
        \STATE Calculate points $X_{1}$, $X_{2}$ .
        \STATE $\hat{s} \gets \mathrm{null}$ 
        \STATE $S_c \gets \mathrm{null}$ 
        \STATE $Q_{e}$ are initialized
        \STATE $x = f_{\chi}(S_{x}, X_{1}, X_{2})$   %\hfill\COMMENT{list of x values that satisfy $x\in R_{1}$}
        \FOR {$x_{i}$ in $x$}
        %\hfill\COMMENT{$s\in R_{1}(i)$}
        \STATE $s_{i,c} = A_{x_i}\left(f_{\psi,i}(A_{i}(:,2), Y_{1}, Y_{2}),2\right)$
        \STATE APPEND($S_c$, $s_{i,c}$)
        \ENDFOR
        \IF{$S_c$ is not null}
        \STATE $\hat{s} = \arg \min_{s\in S_c}\hspace{4px}\lvert r-s\rvert^2$
        \ENDIF
        \IF{$S_c$ is null}
        \STATE $e \gets$  quadrant of $r$
        \STATE $\hat{s} = \arg \min_{s\in Q_{e}}\hspace{4px}\lvert r-s\rvert^2$
        \ENDIF
        \RETURN $\hat{s}$
    \end{algorithmic}
    \label{algo}
\end{algorithm}
\vspace{-1cm}
\section{Numerical Results}\label{secnum}
In this section, we provide numerical results for the considered network and validate the derived analytical expressions via Monte Carlo simulations with $10^6$ realizations. Specifically, we assume that the transmit power $P_t=10^{-3}$ W, the reference distance $d_0$ is set at 1 m, while $C_0$ and $\sigma_{n}^2$ are set equal to $-30$ dBm and $-140$ dBm, respectively. Moreover, we assume that both BS and CN antennas are omnidirectional, i.e., $G_t=G_r=1$, the path loss exponent $n$ equals to $2.5$, while the distances $d_1$ and $d_2$ are equal to $20$ m and $60$ m. In addition, both CN-RIS and RIS-BS links are assumed to be affected by Nakagami-$m$ fading with shape parameter $m=3$ and scale parameter $\Omega=1$, respectively, whereas $P_{\mathrm{ctr}}$ and $P_{\mathrm{PIN}}$ are given equal to 50 mW and 1 mW.

In Figs. \ref{fig:asep}a and \ref{fig:asep}b, we present the ASEP of HQAM and QAM constellations for $M=64$ and $M=1024$, respectively. As it can be observed, the simulation results validate the tightness of the proposed ASEP approximation of an HQAM constellation, demonstrating the precision of our analysis. Moreover, it can be seen that the adoption of HQAM leads to a reduction in the number of required RIS elements, $N$, in comparison to conventional QAM, which enhances the cost-effectiveness of the system, as fewer reflecting elements are required. Additionally, as $M$ increases, the superiority of HQAM compared to QAM becomes increasingly evident, attributed to the denser symbol allocation on the IQ plane, which enhances spatial efficiency. Furthermore, an increase in the quantization level, $q$, results in a decrease in the ASEP for both modulation schemes, indicating that a higher quantization level can improve network performance. However, the transition from $q=1$ to $q=2$ showcases a more significant performance gain than the progression from $q=2$ to $q=3$ for both $M=64$ and $M=1024$. Therefore, it becomes clear that increasing $q$ beyond 2 is not a practical approach, as it fails to offer significant performance improvements and results in higher power consumption, highlighting the necessity for a cautious selection of $q$ to enhance system performance and efficiency.

Fig. \ref{fig:ee} illustrates the constrained energy efficiency $E_c$ normalized to the network's bandwidth of HQAM and QAM constellations across varying $q$, considering a targeted ASEP of $P_u = 10^{-5}$. Notably, for all the examined $q$ values, HQAM consistently requires a reduced number of reflecting elements to achieve the desired $P_u$, thus highlighting HQAM's contribution to energy efficiency enhancement. Furthermore, despite $q=1$ showing the highest ASEP value for both HQAM and QAM, it is identified as the most favorable option in terms of $E_c$ compared to $q=2$ and $q=3$, attributed to the significant rise in energy consumption that accompanies increasing $q$. Therefore, Fig. \ref{fig:ee} underlines the importance of appropriate selection of $q$ to not only meet performance criteria but also to minimize energy consumption, thus emphasizing the pivotal role of HQAM in achieving enhanced energy efficiency.

\begin{figure}
    \centering
    \begin{tikzpicture}
        \begin{semilogyaxis}[
            width=0.85\linewidth,
            xlabel = $\gamma_r$ (dB),
            ylabel = $P_s$,
            xmin = 0,
            xmax = 40,
            ymin = 0.0001,
            ymax = 1,
            xtick = {0, 5,...,40},
            grid = major,
            legend cell align = {left},
            legend pos = south west,
            legend style={font=\scriptsize}
            ]

             \addplot[
           only marks,
            mark = diamond*,
            mark repeat = 2,
            mark size = 2.5,
            line width = 1pt,
	    style = solid,
            ]
            table {data/detection/det64.dat};
            \addlegendentry{64-HQAM (MLD)}
            \addplot[
            only marks,
            mark = triangle*,
            mark repeat = 2,
            mark size = 2.5,
            line width = 1pt,
	    style = solid,
            ]
            table {data/detection/det256.dat};
            \addlegendentry{MLD 256-HQAM (MLD)}
            \addplot[
            only marks,
            mark = square*,
            mark repeat = 2,
            mark size = 2.5,
            line width = 1pt,
	    style = solid,
            ]
            table {data/detection/det1024.dat};
            \addlegendentry{1024-HQAM (MLD)}

            \addplot[
            mark repeat = 2,
            mark size = 2,
            line width = 1pt,
	    style = solid,
            ]
            table {data/detection/det64.dat};
            \addlegendentry{Proposed Algorithm}
             \addplot[
            mark repeat = 2,
            mark size = 2.5,
            line width = 1pt,
	    style = solid,
            ]
            table {data/detection/det256.dat};

            \addplot[
            no marks,
            mark repeat = 2,
            mark size = 2.5,
            line width = 1pt,
	    style = solid,
            ]
            table {data/detection/det1024.dat};

        \end{semilogyaxis}
    \end{tikzpicture}
    \caption{Detected symbol error rate versus received SNR}
    \label{detection}
\end{figure}
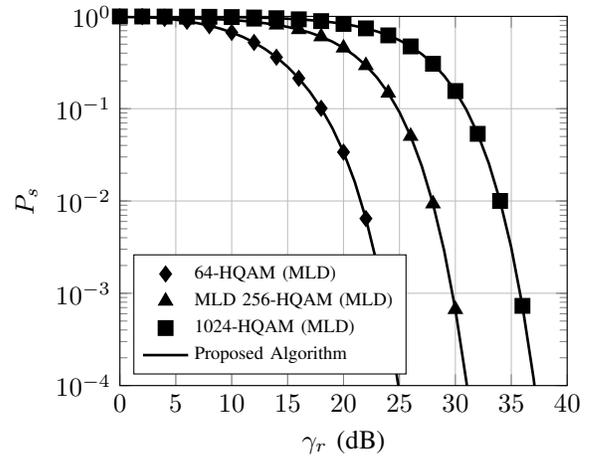

Finally, in Fig. \ref{detection}, we show the detection accuracy of the proposed detection algorithm in comparison to the conventional MLD. Notably, the error rates in detection for the proposed approach are similar to those of MLD, thus our detection algorithm provides the same level of accuracy as MLD, while significantly reducing time complexity from $\mathcal{O}(M)$ to $\mathcal{O}(1)$. Consequently, our algorithm can assist in cases of large $M$, where the complexity of conventional MLD increases linearly with $M$, whereas the complexity of our algorithm remains constant regardless of the value of $M$.

\begin{figure}[h!]
    \centering
    \begin{minipage}{.5\textwidth}
        \centering
\begin{tikzpicture}
	\begin{semilogyaxis}[
	width=0.85\linewidth,
	xlabel = {$N$},
	ylabel = {$P_a$},
	xmin = 50,xmax = 300,
	ymin = 0.00001,
	ymax = 1,
	xtick = {50,100,...,300},
	grid = major,
      legend style = {font = \scriptsize},
	legend cell align = {left},
	legend pos = north east
	]
	\addplot[
	black,
      only marks,
	mark=square*,
        mark repeat=2,
	mark size = 2,
	]
	table {data/ASEP_HQAM_64_q1_.txt};
	\addlegendentry{$q=1$ (Sim.)}
	\addplot[
	blue,
      only marks,
	mark=diamond*,
        mark repeat = 2,
	mark size = 3,
	]
	table {data/ASEP_HQAM_64_q2_.txt};
	\addlegendentry{$q=2$ (Sim.)}
	\addplot[
	brown,
      only marks,
	mark=triangle*,
        mark repeat=2.
	mark size = 3,
	]
	table {data/ASEP_HQAM_64_q3_.txt};
	\addlegendentry{$q=3$ (Sim.)}
	\addplot[
	black,
       no marks,
	line width = 1pt,
	style = solid,
	]
	table {data/ASEP_HQAM_64_q1__theory.txt};	
	\addlegendentry{HQAM (Analytical)}

        \addplot[
	black,
      no marks,
	line width = 1pt,
	style = dashed,
	]
	table {data/ASEP_QAM_64_q1__smoothed.txt};
        \addlegendentry{QAM}
 
	\addplot[
	blue,
       no marks,
	line width = 1pt,
	style = solid,
	]
	table {data/ASEP_HQAM_64_q2__theory.txt};	
	
	\addplot[
	brown,
       no marks,
	line width = 1pt,
	style = solid,
	]
	table {data/ASEP_HQAM_64_q3__theory.txt};

	\addplot[
	blue,
      no marks,
	line width = 1pt,
	style = dashed,
	]
	table {data/ASEP_QAM_64_q2__smoothed.txt};
	\addplot[
	brown,
      no marks,
	line width = 1pt,
	style = dashed,
	]
	table {data/ASEP_QAM_64_q3__smoothed.txt};	
        
	\end{semilogyaxis}
	\end{tikzpicture}
        \subcaption{$M=64$}
        %\subcaption{$M=64$}
        \label{fig:M64}
    \end{minipage}%

\begin{minipage}{.5\textwidth}
        \centering
\begin{tikzpicture}
	\begin{semilogyaxis}[
	width=0.85\linewidth,
	xlabel = {$N$},
	ylabel = {$P_a$},
	xmin = 100,xmax = 1100,
	ymin = 0.00001,
	ymax = 1,
	xtick = {100,300,...,1100},
	grid = major,
      legend style = {font = \scriptsize},
	legend cell align = {left},
	legend pos = north east
	]
	\addplot[
	black,
      only marks,
	mark=square*,
        mark repeat=2,
	mark size = 2,
	]
	table {data/ASEP_HQAM_1024_q1_.txt};
	\addlegendentry{$q=1$ (Sim.)}
	\addplot[
	blue,
      only marks,
	mark=diamond*,
        mark repeat = 2,
	mark size = 3,
	]
	table {data/ASEP_HQAM_1024_q2_.txt};
	\addlegendentry{$q=2$ (Sim.)}
	\addplot[
	brown,
      only marks,
	mark=triangle*,
        mark repeat=2.
	mark size = 3,
	]
	table {data/ASEP_HQAM_1024_q3_.txt};
	\addlegendentry{$q=3$ (Sim.)}
	\addplot[
	black,
       no marks,
	line width = 1pt,
	style = solid,
	]
	table {data/ASEP_HQAM_1024_q1__theory.txt};	
	\addlegendentry{HQAM (Analytical)}

        \addplot[
	black,
      no marks,
	line width = 1pt,
	style = dashed,
	]
	table {data/ASEP_QAM_1024_q1__smoothed.txt};
        \addlegendentry{QAM}
 
	\addplot[
	blue,
       no marks,
	line width = 1pt,
	style = solid,
	]
	table {data/ASEP_HQAM_1024_q2__theory.txt};	
	
	\addplot[
	brown,
       no marks,
	line width = 1pt,
	style = solid,
	]
	table {data/ASEP_HQAM_1024_q3__theory.txt};

	\addplot[
	blue,
      no marks,
	line width = 1pt,
	style = dashed,
	]
	table {data/ASEP_QAM_1024_q2__smoothed.txt};
	\addplot[
	brown,
        no marks,
	line width = 1pt,
	style = dashed,
	]
	table {data/ASEP_QAM_1024_q3__smoothed.txt};	
        
	\end{semilogyaxis}
	\end{tikzpicture}
        %\subcaption{$M=1024$}
        \label{fig:M1024}
        \subcaption{$M=1024$}
    \end{minipage}
    \caption{ASEP versus number of elements}
    \label{fig:asep}
\end{figure}
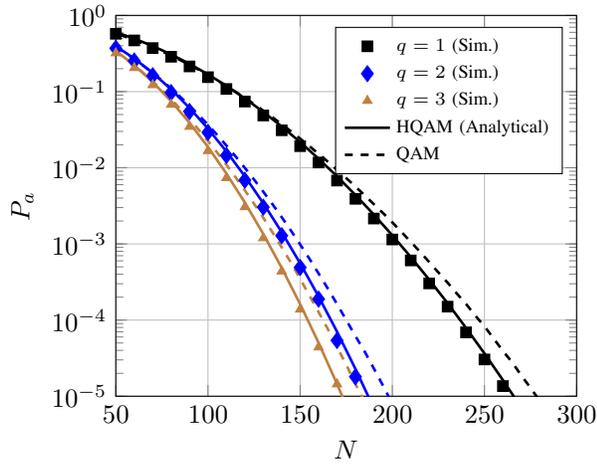
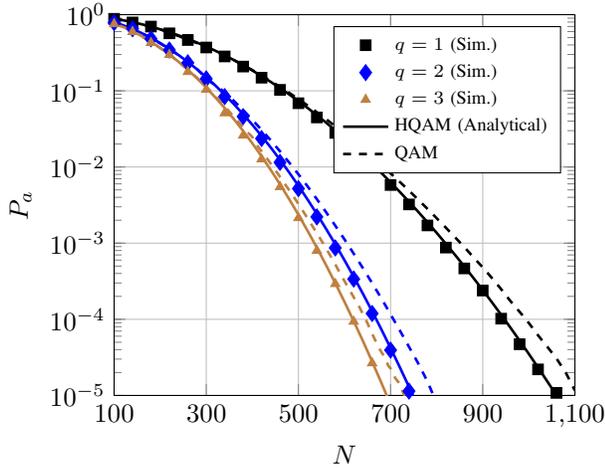

\begin{figure}
    \centering
    \begin{tikzpicture}
        \begin{axis}[
            width=0.85\linewidth,
    	xlabel = {$N$},
    	ylabel = {$E_c/B$ [bits/J/Hz]},
    	xmin = 500,xmax = 1500,
    	ymin = 0,
    	ymax = 12,
    	ytick = {0,2,...,12},
    	xtick = {100,300,...,1700},
    	grid = major,
          legend style = {font = \scriptsize},
    	legend cell align = {left},
    	legend pos = north west
            ]
            
        \addplot[
	black,
        only marks,
	mark=square*,
        mark repeat=7,
	mark size = 2,
	]
	table {data/EE_HQAM_1024_q1__final2.txt};
	\addlegendentry{$q=1$ (Sim.)}
	\addplot[
	blue,
        only marks,
	mark=diamond*,
        mark repeat = 7,
	mark size = 2,
	]
	table {data/EE_HQAM_1024_q2__final2.txt};
	\addlegendentry{$q=2$ (Sim.)}
	\addplot[
	brown,
        only marks,
	mark=triangle*,
        mark repeat=7,
	mark size = 2,
	]
	table {data/EE_HQAM_1024_q3__final2.txt};
	\addlegendentry{$q=3$ (Sim.)}
	\addplot[
	black,
        no marks,
	line width = 1pt,
	style = solid,
	]
	table {data/EE_HQAM_1024_q1__final2.txt};	
	\addlegendentry{HQAM}

        \addplot[
	black,
        no marks,
	line width = 1pt,
	style = dashed,
	]
	table {data/EE_QAM_1024_q1__final2.txt};
        \addlegendentry{QAM}
 
	\addplot[
	blue,
        no marks,
	line width = 1pt,
	style = solid,
	]
	table {data/EE_HQAM_1024_q2__final2.txt};	
	
	\addplot[
	brown,
        no marks,
	line width = 1pt,
	style = solid,
	]
	table {data/EE_HQAM_1024_q3__final2.txt};

	\addplot[
	blue,
        no marks,
	line width = 1pt,
	style = dashed,
	]
	table {data/EE_QAM_1024_q2__final2.txt};
	\addplot[
	brown,
        no marks,
	line width = 1pt,
	style = dashed,
	]
	table {data/EE_QAM_1024_q3__final2.txt};
        \end{axis}
    \end{tikzpicture}
    \caption{Normalized $E_c$ versus $N$ for $M=1024$}
    \label{fig:ee}
\end{figure}
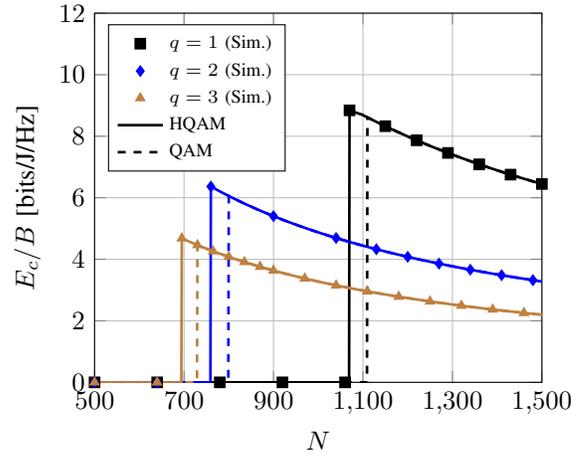

\vspace{-3mm}
\section{Conclusions}\label{conclusion}
\vspace{-3mm}
In this work, we explored the application of HQAM in RIS-assisted networks, focusing on its potential to minimize the number of reflecting elements. Specifically, we developed analytical expressions for the ASEP and introduced a novel metric for conditioned energy efficiency, assessing the network's energy efficiency while maintaining ASEP below a threshold value. Furthermore, we propose a novel detection algorithm for HQAM that implements sphere decoding in $\mathcal{O}\left(1\right)$ complexity. Our findings, validated through Monte Carlo simulations, highlighted HQAM's superiority in energy efficiency and ASEP over traditional QAM, offering significant insights for future wireless systems.

\section*{Acknowledgment}
% This research has received funding from the European Union’s Horizon Europe Framework Programme under grant agreement No 101096456.
This work was funded from the Smart Networks and Services Joint Undertaking (SNS JU) under European Union's Horizon Europe research and innovation programme (Grant Agreement No. 101096456 - NANCY). 

\bibliographystyle{IEEEtran}
\bibliography{bibliography}
\end{document}